\documentclass{mem}
\usepackage{natbib}
\usepackage{graphicx}
\idline{75}{282}
\begin{document}
\def\teff{$T\rm_{eff }$}
\def\kms{$\mathrm {km s}^{-1}$}

\title{
Pattern Speed of Lopsidedness in Galactic Disks
}

\author{
Chanda \,J. \,Jog\inst{1} 
          }

  \offprints{C.J. Jog}

\institute{
Department of Physics, Indian Institute of Science, Bangalore 560012, India $\: \: \:$
\email{cjjog@physics.iisc.ernet.in}
}

\authorrunning{Jog }

\titlerunning{Lopsidedness in galactic disks}

\abstract{
The disks of spiral galaxies commonly show a lopsided mass distribution, with a typical 
fractional amplitude of 10\% for the Fourier component m=1. This is seen in both stars and gas, and
the amplitude is higher by a factor of two for galaxies in a group. The study of lopsidedness is a new topic, in contrast to the extensively studied bars and two-armed spirals (m=2).
Here, first a brief
overview of the observations of disk lopsidedness is given, followed by a summary of
the various mechanisms that have been proposed to explain its physical origin.
These include tidal interactions, gas accretion, and a global instability. The pattern speed of lopsidedness 
in a real galaxy
has not been measured so far, the various issues involved will be discussed. Theoretical studies have
 shown that 
the m=1 slow modes are long-lived, while the modes with a moderate pattern speed
 as triggered in interactions,
last for only about a Gyr. Thus a measurement of the 
pattern speed of lopsided distribution will help 
identify the mechanism for its origin. 
}
\maketitle{}

\section{Introduction}

It is known that the light and hence the mass
distribution in disks of spiral galaxies is not strictly
axisymmetric, as for example in M101 or in NGC 1637,
where the isophotes are elongated in one half of the galaxy. 
This phenomenon was first highlighted in the paper by Baldwin, Lynden-Bell, \& Sancisi (1980) 
for the atomic hydrogen gas in the outer regions in the two  halves of some
galaxies, and they called these  `lopsided' galaxies. A galaxy is said to be lopsided if it displays a non-axisymmetric
mass distribution of type $m=1$ where $m$ is the azimuthal wavenumber, or a cos$\phi$ distribution where $\phi$ is the azimuthal angle in the plane of the disk.
From the HI global velocity profiles compiled for a large sample of galaxies, Richter \& Sancisi (1994) concluded that nearly half the galaxies show lopsidedness.

\begin{figure}[h]
\centering
\includegraphics[height=1.7in,width=2.0in]{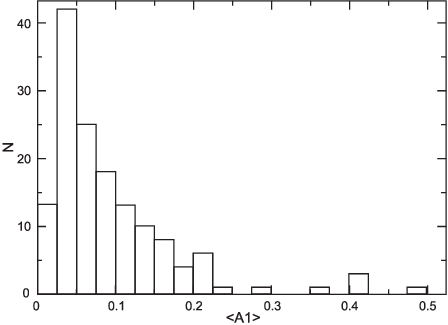}
\caption{\footnotesize The histogram showing the number of galaxies  vs. the lopsided amplitude, measured in 149 galaxies at the inclination of $< 70^0$ from the OSU sample (Bournaud et al. 2005). The typical normalized lopsided amplitude $<A_1>$ measured over the radial range between 1.5-2.5 disk scalelengths is = 0.1. Thus most spiral galaxies show significant lopsidedness.} 
 \label{fig1}
\end{figure}

The near-IR observations show that lopsidedness is common in stars as well. The stellar disks in 
30 \% of galaxies studied are significantly lopsided, with an amplitude A$_1$ $> 0.1 $. This is measured as the Fourier amplitude of the m=1 component normalized to the average value (Rix \& Zaritsky 1995). The lopsidedness is stronger at larger radii. This was confirmed for a larger sample of 149 galaxies from the Ohio State University  (OSU) database by Bournaud et al. (2005), see Fig. 1 here.
Thus, lopsidedness is a typical feature, hence it is important to study its dynamics.

\medskip

In a first such study, a similar Fourier analysis for the interstellar atomic hydrogen gas (HI) has been done by analyzing the two-dimensional surface density plots for 18 galaxies in the Eridanus group (Angiras et al. 2006).
Using HI as the tracer allows the asymmetry to be measured upto several times the disk scalelengths, see Fig. 2. This is more than twice the radial distance covered for the stars, 
since the sky background in the near-IR limits the Fourier analysis to $\sim 2.5$ disk scalelengths. 
Note that, during the Fourier decomposition, the same centre has to be used for all the annular radial bins (Rix \& Zaritsky 1995,
Jog \& Combes 2008).

\begin{figure}[h]
\centering
\includegraphics[height=1.25in,width=2.5in]{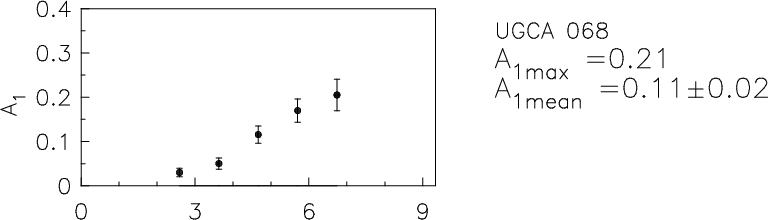}
\medskip 
\includegraphics[height=1.25in,width=2.5in]{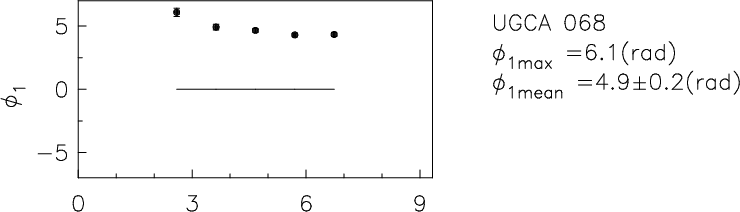} 
\caption{\footnotesize The lopsided amplitude and phase of the HI surface density distribution 
vs. radius (in units of the disk scalelength) for the galaxy UGC 068 in the Eridanus group, taken
from Angiras et al. (2006). The amplitude increases with radius (Fig. 2a), and the 
phase is nearly constant indicating that m=1 is a global mode (Fig. 2b).} 
\end{figure}

\bigskip

\begin{figure}[h]
\centering
\includegraphics[height=1.8in,width=2.0in]{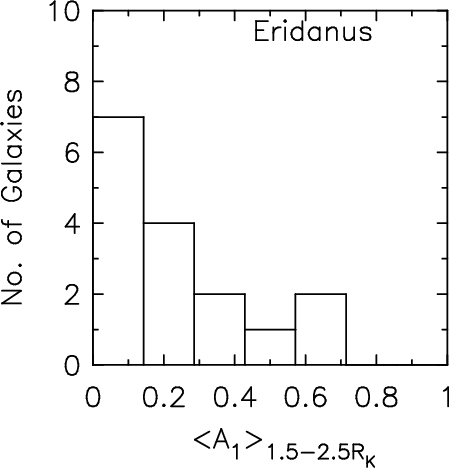}
\caption{\footnotesize The histogram showing the number of galaxies vs. the lopsided amplitude, A$_1$, measured in the 1.5-2.5 disk scalelength range for the Eridanus group galaxies, from Angiras et al. (2006). The group galaxies are more lopsided,  compare with Fig. 1.} 
\end{figure}

Since these galaxies belong to a group, this serendipitously allows a study of the effect of group environment on lopsidedness.
The average lopsided amplitude A$_1$ for the group galaxies is higher $\sim 0.2$,
see the histogram in Fig. 3. The fraction of galaxies showing lopsidedness is higher in groups. For example, all the group galaxies have A$_1$ $> 0.1$, which is the mean value for the field case. The frequent interactions in galaxies in groups could explain the higher
  fraction of galaxies showing lopsidedness.

\section {Origin of disk lopsidedness}

A non-axisymmetric feature will tend to get wound up in a few dynamical timescales or  in 
 a few $\times$
 10$^8$ yr, due to the differential rotation in the galactic disk. This is the well-known winding dilemma. However, a high fraction of galaxies is observed to be lopsided. To get around the limited lifetime implied by the winding problem, a number of models have been proposed.

\subsection {Kinematical model}
Starting with aligned orbits, Baldwin et al. (1980) calculated  the winding up time while taking account of the epicyclic motion of the particles. This decreases the effective radial range over which the differential rotation is effective, hence it increases the lifetime to be about 5
 times longer than for the usual material arms, or $\sim$ a Gyr.
This is still much smaller than the Hubble time so they concluded that the lopsidedness seen could not be of primordial origin.

\subsection {Dynamical models} 

The various mechanisms that have been proposed to explain the physical origin of the disk lopsidedness are discussed next.

\subsubsection {Disk response to lopsided potential}

The basic features of the orbits, isophotes and kinematics in a lopsided galaxy can be understood  by treating the disk response to an imposed lopsided perturbation potential (Jog 1997, Schoenmakers et al. 1997). The origin of lopsided halo potential was
assumed to be of tidal origin. The results from Jog (1997, 2000) are summarized below:

The unperturbed axisymmetric potential, $\psi_0$, and the first-order perturbation
 potential, $\psi_{lop}$, are taken to be 
respectively:

$$ \psi_0 = V_c^2 \: ln R   \eqno(1) $$

$$ \psi_{lop} = V_c^2 \: {\epsilon_{lop}}  cos {\phi}  \eqno(2)$$

\noindent where V$_c$ is the rotation velocity in the region of flat rotation, and $\epsilon_{lop}$ denotes the small
perturbation parameter in the potential which is taken to be constant with radius. The perturbation potential is taken to be non-rotating for simplicity.

The equations of motion are solved using the first order epicyclic theory. The perturbed closed loop orbits are given by:

$$ R = R_0 ( 1 - 2 {\epsilon_{lop}} cos {\phi} ) \eqno (3) $$
$$ V_R = 2 V_c  \: {\epsilon_{lop}} sin {\phi}  \eqno (4) $$
$$ V_{\phi} = V_c ( 1 + {\epsilon_{lop}} cos {\phi} ) \eqno (5) $$

Thus an orbit is elongated along $\phi$ = 180$^0$ or the minimum of the perturbation potential, and it is shortened along the opposite direction (along $\phi$ = 0$^0$).

The observations measure isophotes rather than orbits, hence the isophotal shapes are obtained for an exponential galactic disk in a lopsided potential. {\it The resulting isophotes have an egg-shaped oval appearance}. This agrees with the observed isophotal shapes in M101. On solving the equations of perturbed motion (eqs.[3]-[5]), the equation of continuity, and the effective surface density together, one gets:

$$ {\epsilon_{lop}} = {\frac {A_1}{(\frac{2 R}{R_{exp}}) - 1}}   \eqno (6)$$

\noindent From this, the typical value of the perturbation potential parameter $\epsilon_{lop}$ is obtained to be
$\sim 0.05$ for the average observed lopsided amplitude.
This denotes the global distortion of the halo potential. Thus, the disk lopsidedness can be used as a diagnostic to study the lopsidedness of the halo potential.

The disk response to a distorted halo that gives spatial lopsidedness also results in kinematical lopsidedness in the disk. The resulting rotation curve is asymmetric (eq.[5]).
The rotational velocity is maximum along $\phi = 0^0$ and minimum along the opposite direction. The typical difference in the two halves of a galaxy is $ 2 \epsilon_{lop} V_c$ or $\sim$ 10 \% of the rotation velocity,
 $\sim$ 25 km s$^{-1}$ (Jog 2002).  Thus the observers should give the full two-dimensional data
for the azimuthal velocity, rather than an azimuthally averaged value. The observed asymmetry in the rotation curve then can be used as a tracer to determine the magnitude of the lopsided potential.

\subsubsection {Tidal interactions}

Tidal interactions between galaxies have long been suggested as a possible mechanism for the origin of disk
lopsidedness. The magnitude and age of lopsidedness thus generated have been studied by N-body simulations by Bournaud et al. (2005) for various mass ratios of the galaxies, orbits, and inclinations.
 The simulations include stars, gas and the dark matter halo, and all three are taken to be live. 

The lopsidedness  generated in a prograde encounter between two spiral galaxies with a mass ratio 2:1 is shown in Fig. 4.
A strong lopsidedness is triggered during the interaction, but it lasts for only $\sim 2$ Gyr. 
Thus this mechanism cannot explain the high A$_1$ seen in many isolated galaxies. The latter may be explained due to an interaction with or accretion of a satellite with a much smaller mass, but this can also cause a vertical puffing up of the disk. This idea has to be explored systematically.
In a recent paper, Mapelli, Moore \& Bland-Hawthorn (2008) show that a fly-by encounter with a neighbouring galaxy UGC 1807 is the preferred mechanism for the origin of lopsidedness seen in NGC 891.

\begin{figure}[h]
\centering
\includegraphics[height=3.0in,width=2.2in]{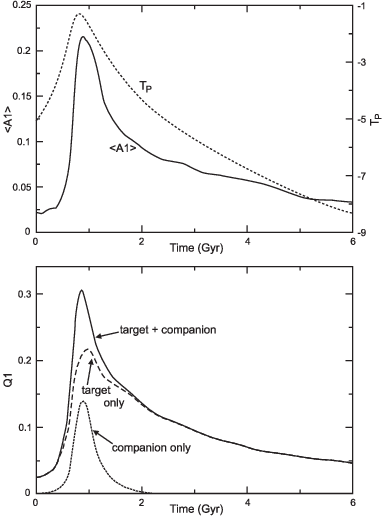}
\caption{\footnotesize Plot of A$_1$ vs. time in the middle radial range of 1.5-2.5 disk scalelengths, generated in a distant interaction between galaxies (Bournaud et al. 2005). 
The peak value is $\sim 0.2$, higher than the average value observed for the field case, but it drops rapidly to 0.05 in a few Gyr. 
The lower panel shows the same in terms of Q$_1$, the cumulative potential from the disk.
} 
\end{figure}

Tidal interactions appear to be the dominant mechanism for the
 generation of lopsidedness in the group galaxies. First, tidal interactions are frequent in a group because of the high number density of galaxies.
 Second, the early type galaxies show a higher lopsidedness in group galaxies, as would be
expected if lopsidedness arises due to tidal encounters (Angiras et al. 2006).

\subsubsection {Gas accretion}

\begin{figure}[h]
\centering
\includegraphics[height=1.4in,width=2.65in]{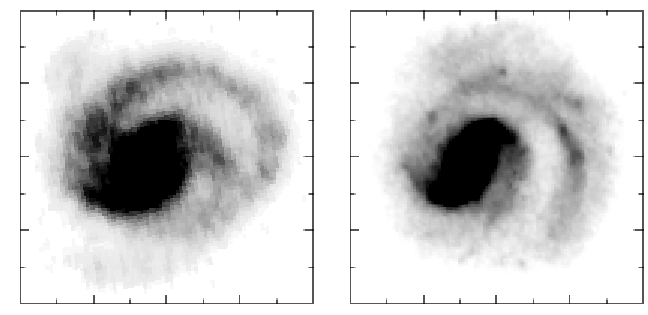}
\caption{\footnotesize NGC 1637 - the lopsidedness due to gas accretion (l.h.s.) matches the near-IR data (r.h.s.),
from Bournaud et al. (2005).} 
\end{figure}

External gas accretion onto a galaxy from a cosmological filament can give rise 
to disk lopsidedness, as shown by Bournaud et al. (2005). This can explain the lopsidedness seen in many isolated galaxies.
They used reasonable values of accretion rates of 4 and 2 M$_{\odot}$ yr$^{-1}$ on two sides of a disk, which
would double the galaxy mass in a Hubble time. The accretion
was shown to explain the main features including the lopsidedness observed in
 the near-IR map of NGC 1637, see Fig. 5. Here once the accretion stops, the lopsidedness disappears within $\sim 2$ Gyr.

\subsubsection {Global m=1 instability}

An internal mechanism involving a global mode instability as the generation mechanism was explored by Saha, Combes \& Jog (2007). They treated a slowly rotating, global m=1 mode in a purely exponential galactic disk, and showed that the inclusion of self-gravity 
makes it long-lasting. This model was 
motivated by the nearly constant phase of lopsidedness that is observed (see e.g., Fig. 2), 
which indicates a global mode.
A unique feature of the m=1 mode is that it
shifts the centre of mass of the disturbed galaxy away from the original centre of mass. 
This further acts as an indirect force on the original centre of mass, which results in long-lived m=1 modes. 

Using the linearized fluid equations
and the softened self-gravity of the perturbation, a self-consistent
quadratic eigenvalue equation was derived and solved.
 The resulting mode gives the basic observed feature, namely the fractional Fourier amplitude A$_1$ increases with radius. 
Fig. 6 shows the resulting isodensity contours, clearly the centres of isocontours are
progressively more disturbed in the outer parts.

\begin{figure}[h]
\centering
{\rotatebox{270}{\includegraphics[height=2.0in,width=2.0in]{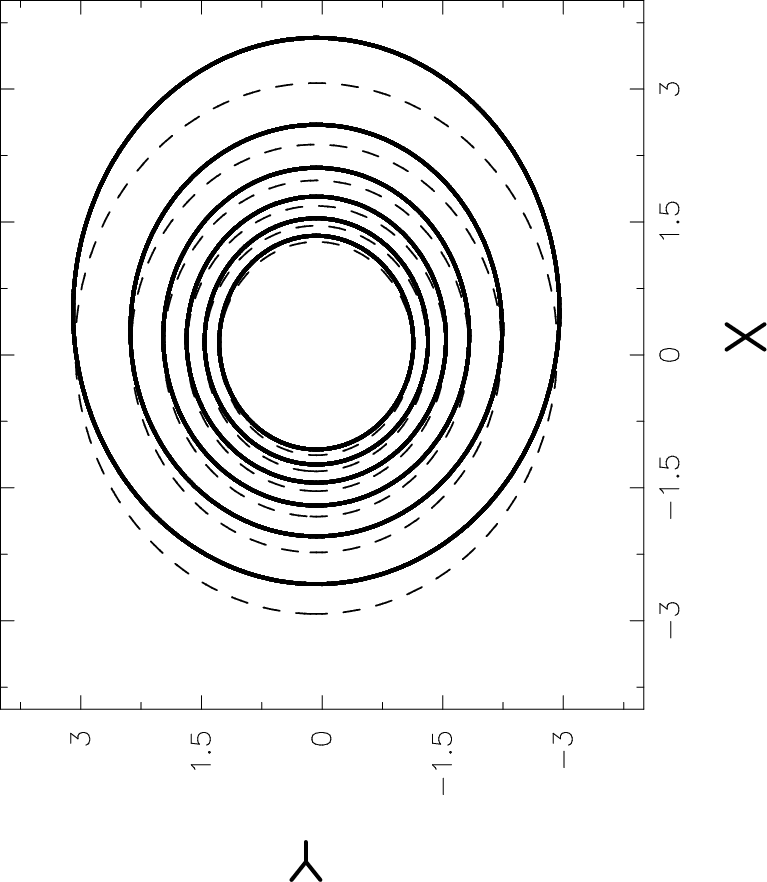}}}
\caption {Contours of constant surface density for a global m=1 mode, from Saha et al. (2007). Here the x and y axes
are given in units of the disk scalelength. The maximum surface density occurs at (0,0). The outer
contours show a progressive deviation from the undisturbed circular distribution, indicating
a more lopsided distribution in the outer parts- as observed. }
\end{figure}

\begin{figure}[h]
\centering
{\rotatebox{270}{\includegraphics[height=2.0in,width=2.0in]{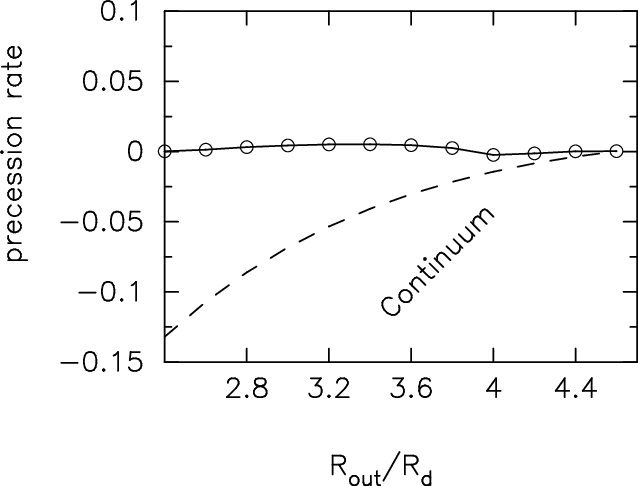}}} 
\caption{The precession rate for the global lopsided mode vs. the size of the disk in units of the disk scalelength
in a galactic disk (shown as the line with circles), from Saha et al. (2007). The precession rate 
is very low, thus the mode is long-lived. The dashed line denotes the free precession rate $(\kappa - \Omega)$.
}
\end{figure}

The self-gravity of the mode results in a significant
reduction in the differential precession, by a factor of $\sim 10$ compared to the free
precession. This leads to 
persistent m=1 modes, as shown in Fig. 7. The lifetime or the winding up time is large $\sim 12$ Gyr.
These results are confirmed by N-body simulations. 
Future work on this topic should include a live halo to get a more realistic picture. 

\section{Pattern speed of lopsidedness}

The pattern speed of lopsided distribution has not been well-studied so far. In fact its value in a real galaxy has not been measured yet, although this would be important to study. First, a rotating lopsided pattern involves rich physics and will led to secular evolution, similar to the case of a rotating bar (m=2). Second, the value of pattern speed can  help 
identify the main mechanism for the origin of lopsidedness as argued next.

Generally the pattern speed of lopsided distribution is assumed to be zero for simplicity (Rix \& Zaritsky 1995, Jog 1997). Theoretical studies of slow m=1 modes show these to be long-lived:  as in the near-Keplerian central region in M31 
(Tremaine 2001), and for
 the global modes in a pure exponential  galactic disk (Saha et al. 2007), and for
 the m=1 modes arising due to swing amplification in the outer parts of a galactic disk in a rigid halo (Dury et al. 2008).
Note that such slow m=1 modes are appealing since these are long-lived and hence can explain lopsidedness seen in isolated galaxies.

The theoretical work so far shows that the pattern speed has a crucial effect on the lifetime of the lopsided distribution. The global m=1 modes in a pure disk are long-lived, lasting for nearly a Hubble time as shown by analytical as well as N-body simulations (Saha et al. 2007). On the other hand, the m=1 modes with a non-zero, moderate pattern speed 
in a disk plus halo system
%generated in a disk that is in the field of a dark matter halo 
last for only $\sim$ 1-2 Gyr, as shown in an important paper by Ideta (2002). The short life is attributed to the wake
generated in the halo due to the m=1 mode in the disk. Such a short lifetime of lopsidedness  
is confirmed in simulations of tidal interactions and satellite mergers that include both the disk and a live halo (Bournaud et al. 2005).
The detailed dynamics for this process needs to be explored more in future simulations.

The tidal picture of origin of lopsidedness is expected to give rise to an m=1 pattern with a moderate pattern speed.
This speed is given by the relative velocity over the impact parameter (Ideta 2002), hence has a typical value of a few km s$^{-1}$ for field galaxies. Such modes are short-lived as discussed above.
 This mechanism may be dominant in the group galaxies (Angiras et al. 2006, Angiras et al. 2007), as argued in Section 2. 
Despite the short life-time expected for such modes,  the high frequency of lopsided galaxies observed in a group could be 
explained by
the repeated tidal interactions that are expected to occur between galaxies in a group.
The origin due to gas accretion, on the other hand, may give rise to slow modes though this has to be checked. 

Thus the measurement of pattern speed of lopsided distribution in a real galaxy is necessary. This will shed light on the mechanism for its origin and its lifetime.

\subsection {Measurement of pattern speed}

The various issues involved with the measurement of pattern speed of lopsidedness (m=1) in a galactic disk are discussed next.
The pattern speed for the bars and the two-armed spirals (m=2) has been studied extensively in the literature,
this case has also been discussed in several articles in these proceedings.
 In contrast, the measurement of  pattern speed for the m=1 distribution 
is an open topic.

  In the case of m=2, the resonance points denoting the Inner Lindblad Resonance (ILR) and 
the corotation (CR), along with a knowledge of the rotation curve, are used to get limits on the value of the pattern speed of the bar. However, for m=1, a similar constraint cannot be applied. For m=1, the ILR is given by:
 
$${\Omega}_p   =  \Omega  -  \kappa    \eqno (7) $$ 

This equation cannot be satisfied when $\Omega_p > 0$  since
$\kappa > \Omega$ for a self-gravitating disk. Thus the ILR condition for m=1 (eq.[7]) is only valid for a negative pattern speed. In itself there is no problem with this. In fact,  the retrograde encounters more readily give a lopsided distribution as seen from numerical simulations (Bournaud et al. 2005),
 and the pattern speed for this is likely to be negative.
However, when $\Omega_p < 0$, then the condition for CR (namely, $\Omega = \Omega_p$) is not satisfied physically. Thus one cannot define both ILR and CR for a given value of the pattern speed for m=1. Hence the latter cannot be determined using the resonance points.

The other possible ways to determine the pattern speed of the lopsided distribution are as follows:

\begin{itemize}
\item   The residual velocity fields could be used to determine $\Omega_p$, similar to what
was done for m=2 by Canzian (1993). 
\item A kinematical method, such as the one proposed by  Tremaine \& Weinberg (1984), could be applied for the m=1 case. 
\item For m=2, the star formation ring in the disk is used as a tracer
to determine the location of the ILR. This may not be applicable for m=1 since the molecular gas which forms the site of star formation lies within the inner two disk scalelengths, while the lopsidedness 
is mostly seen in the outer parts (Jog 1997). Hence
the molecular gas is not a good tracer of lopsidedness.
\end{itemize}

Thus the measurement of pattern speed of lopsidedness in a galactic disk remains an open problem.

\section{Conclusions}
Lopsidedness is a common feature of spiral galaxies, seen in both stars and gas, and in the field as well as the group galaxies. The lopsidedness has an amplitude of $\sim 10$ \% in the radial range of 1.5-2.5 disk scalelengths, and is higher at larger radii.

Various physical mechanisms have been
proposed to explain the origin of lopsidedness. These  include
external ones such as tidal interactions and gas accretion, or an internal one such as a global m=1 instability.
These could all play a role to a varying degree, and which particular mechanism dominates depends on the specifics of the case. For example, in group galaxies, the tidal interactions seem to play a dominant role in generating lopsidedness.

The pattern speed of lopsidedness is 
 an important parameter, but it has not yet been measured in any galaxy.

For a recent review of lopsided spiral galaxies, see Jog \& Combes (2008).

\bigskip

\begin{acknowledgements}
It is a pleasure to thank Victor Debattista and Enrico Maria Corsini for inviting me to this stimulating workshop. 
\end{acknowledgements}

\bigskip

\noindent{\bf References}

\noindent    Angiras, R.A., Jog, C.J., Omar, A., \& Dwarakanath, K.S.
  2006, MNRAS, 369, 1849

\noindent Angiras, R.A., Jog, C.J., Dwarakanath, K. S., \&
  Verheijen, M.A.W. 2007, MNRAS, 378, 276

\noindent  Baldwin, J.E., Lynden-Bell, D., \& Sancisi, R. 1980,
  MNRAS, 193, 313

\noindent Bournaud, F., Combes, F., Jog, C.J., \& Puerari, I.
 2005 , A \& A , 438, 507

\noindent Canzian, B. 1993, ApJ, 414, 487

\noindent Dury, V., De Rijcke, S., Debattista, V.P., \&  Dejonghe, H. 2008, MNRAS, 387, 2

\noindent Ideta, M. 2002, ApJ, 568, 190

\noindent  Jog, C.J. 1997, ApJ, 488, 642

\noindent  Jog, C.J. 2000, ApJ, 542, 216

\noindent  Jog, C.J. 2002, A \& A, 391, 471

\noindent Jog, C.J., \& Combes, F. 2008, Physics Reports, submitted (arxiv.org: 0811.1101)

\noindent Mapelli, M., Moore, B., \& Bland-Hawthorn 2008, MNRAS,  388, 697

\noindent Richter, O.-G., \& Sancisi, R. 1994, A \& A, 290, L9 

\noindent Rix, H.-W., \& Zaritsky, D. 1995, ApJ, 447, 82

\noindent Saha, K., Combes, F., \& Jog, C.J.  2007, MNRAS, 382, 419

\noindent  Schoenmakers, R.H.M., Franx, M., \& de Zeeuw, P.T.
   1997, MNRAS, 292, 349

\noindent Tremaine, S. 2001, AJ, 121, 1776

\noindent Tremaine, S., \& Weinberg, M. 1984, ApJ, 282, L5

\end{document}